\documentclass[aps,prb,twocolumn,showpacs,superscriptaddress]{revtex4}

\usepackage{graphicx}

\usepackage{graphicx}
\usepackage{color}
\usepackage[normalem]{ulem}

\newcommand{\bise}{Bi$_2$Se$_3$}
\newcommand{\bite}{Bi$_2$Te$_3$}
\newcommand{\sbte}{Sb$_2$Te$_3$}
\newcommand{\kv}{\textbf{k}}
\newcommand{\abinitio}{\textit{ab initio}}
\newcommand{\gb}{$\bar{\Gamma}$}
\newcommand\myurl[2]{#1}
\providecommand{\eg}{e.g.,}
\providecommand{\etal}{\emph{et al.}}
\providecommand{\ie}{i.e.,}

\begin{document}


\title{Bulk band structure of Bi$_2$Te$_3$}


\author{Matteo Michiardi}
\affiliation{Department of Physics and Astronomy, Interdisciplinary Nanoscience Center, Aarhus University, 8000 Aarhus C, Denmark}
\author{Irene Aguilera}
\affiliation{Peter Gr\"unberg Institute and Institute for Advanced Simulation, Forschungszentrum J\"ulich and JARA, D-52425 J\"ulich, Germany}
\author{Marco Bianchi}
\affiliation{Department of Physics and Astronomy, Interdisciplinary Nanoscience Center, Aarhus University, 8000 Aarhus C, Denmark}
\author{Vagner Eust\'aquio de Carvalho}
\author{Luiz Orlando Ladeira}
\author{Nayara Gomes Teixeira}
\author{Edmar Avellar Soares}
\affiliation{Departamento de F\'isica, ICEx, Universidade Federal de Minas Gerais, 30123-970  Belo Horizonte, MG, Brazil}
\author{Christoph Friedrich}
\author{Stefan Bl\"ugel}
\affiliation{Peter Gr\"unberg Institute and Institute for Advanced Simulation, Forschungszentrum J\"ulich and JARA, D-52425 J\"ulich, Germany}
\author{Philip Hofmann}
\affiliation{Department of Physics and Astronomy, Interdisciplinary Nanoscience Center, Aarhus University, 8000 Aarhus C, Denmark}


\date{\today}

\begin{abstract}
The bulk band structure of Bi$_2$Te$_3$ has been determined by angle-resolved photoemission spectroscopy and 
compared to first-principles calculations. We have performed calculations 
using the local density approximation (LDA) of density functional theory and 
the one-shot $GW$ approximation within the all-electron full-potential linearized augmented-plane-wave 
(FLAPW) formalism, fully taking into account spin-orbit coupling. Quasiparticle effects produce 
significant changes in the band structure of \bite~when 
compared to LDA. Experimental and calculated results are compared in the spectral regions where 
distinct differences between the LDA and $GW$ results are present. Overall a superior agreement 
with $GW$ is found, highlighting the importance of many-body effects in the band structure of this 
family of topological insulators.
\end{abstract}


\maketitle

\section{Introduction}
In the last few years topological insulators have been the subject of extensive studies, both experimental and theoretical.\cite{Moore2009,Hasan2010} For these materials, the peculiarity of having a bulk band parity inversion in an odd number of time-reversal invariant momenta across the Brillouin zone (BZ) gives rise to the existence of topological surface states protected by time-reversal symmetry.\cite{fu2007} The surface states crossing the band gap are necessarily metallic and present spin helicity, a condition where the electrons' spin is locked to their momentum, opening the possibility of applications in spintronics.
 While many topological insulators have been predicted and experimentally investigated, the bismuth chalcogenides are by far the most extensively studied. 
 
For experimental studies, Bi$_2$Se$_3$ and Bi$_2$Te$_3$ are considered to be prototypical materials.\cite{Zhang2009,Xia2009,Chen:2009} They share the same rhombohedral crystal structure, which consists of quintuple layers bound to each other through weak van der Waals forces, giving easy access to the (111) surface by cleavage. Their surface electronic structure is also very similar in the sense that both support a single closed surface Fermi contour around the $\bar{\Gamma}$ point of the (111) surface BZ. In both cases, this is caused by a parity inversion at the bulk $\Gamma$ point. However, while the surface electronic properties are similar, there are some significant differences in the detailed dispersion of the surface states. Both materials show a warping of the surface state Dirac cone far away from the Dirac point but for Bi$_2$Te$_3$ this warping is much more pronounced than for Bi$_2$Se$_3$.\cite{Chen:2009,Fu:2009,Kuroda:2010,Bianchi:2012b} 

The materials also show significant differences in their bulk electronic structure. For instance, as was recently shown via a comparison between angle-resolved photoemission (ARPES) measurements and $GW$ calculations,\cite{Nechaev2013,Yazyev2012} Bi$_2$Se$_3$ has a direct band gap with the bulk valence band maximum (VBM) and conduction band minimum (CBM) located at the $\Gamma$ point of the BZ, whereas the character of the band gap in Bi$_2$Te$_3$ is still subject to debate.\cite{Kioupakis2010,Yavorsky2011,Nechaev2013-2} 

So far, \abinitio~calculations for topological insulators have been mainly performed using either the local density (LDA) or generalized gradient (GGA) approximations of density functional theory (DFT). The reason is that LDA and GGA consitute an efficient approach which allows for the study not only of the bulk but also of the surface states. Since the parity inversion in the bulk states is mainly caused by spin-orbit coupling (SOC) effects, including these proved to be crucial. LDA and GGA calculations of the surface states have mostly shown good agreement with the experimental results.\cite{Zhang2009,Herdt2013} 
However, these two approaches fail to correctly describe some important aspects of the bulk band structures. Vidal \etal\cite{Vidal2011} 
have demonstrated that LDA may incorrectly predict trivial insulators as topological insulators. In addition, several $GW$ studies of topological insulators in the last few 
years\cite{Kioupakis2010,Yazyev2012,Sakuma2011,Vidal2011,Svane2011,Zhu2013,Aguilera2013,Nechaev2013,Nechaev2013-2} 
have shown that the nature of the band gap (indirect or direct), its magnitude, and the effective masses of the 
bands involved in the band inversion are not described correctly within LDA but are often corrected with the $GW$ approximation. 

As an example, LDA predicts Bi$_2$Se$_3$ to be an indirect gap semiconductor with the VBM close to the Z point,\cite{Zhang2009,Yazyev2012} whereas the inclusion of a $GW$ correction correctly reproduces the direct band gap, in agreement with experiment.\cite{Yazyev2012,Nechaev2013} Such details are also crucial for the surface electron dynamics as they determine if the surface states are degenerate with bulk states at another ${\rm k}_{\parallel}$ or not. In fact, reaching the topological transport regime\cite{Hsieh2009} would not be possible for Bi$_2$Se$_3$ as described by LDA but it is possible in the case where the Dirac point lies within the direct band gap at $\bar{\Gamma}$.\cite{Nechaev2013} 

Calculations using the $GW$ approach were also able to reproduce more subtle details in the experimentally observed dispersion. The first $GW$ bulk band structure of Bi$_2$Se$_3$ showed that the band-inversion-induced characteristic M-shaped dip in the band forming the VBM disappears upon the inclusion of $GW$ corrections.\cite{Yazyev2012} Again, this is in good agreement with the results from ARPES for a cut exactly through the $\Gamma$ point but not for a cut at a different ${\rm k}_{\rm \perp}$ value in the $\Gamma$--Z direction (for a good illustration of this, see the supporting material of Ref.~\onlinecite{Bianchi:2010b}). Also using the $GW$ approximation, Nechaev \emph{et al.} were later able to show that the M-shaped dip reappears for larger ${\rm k}_{\rm \perp}$, as seen in the bulk bands projected onto the (111)-surface BZ.\cite{Nechaev2013}

In addition to the \textit{a posteriori} observation that $GW$ is in general in better 
agreement with ARPES experiments than LDA, a comparison between $GW$ and ARPES is also 
more justified from a fundamental point of view. LDA is meant to predict 
ground-state properties and not excitation energies; $GW$, on the other hand, is constructed 
to describe the energies required to remove an electron from or add an electron to the system (the quasiparticle 
energies), in clear correspondence to the excitation energies measured in direct and inverse 
photoemission experiments.

In order to assess the accuracy of the LDA and $GW$ approximations for \bite, we present bulk band
structures of this topological insulator obtained in LDA and $GW$ calculations and compare
them with band structure measurements by ARPES. Particular emphasis is put on the differences between the 
LDA and $GW$ results in the regions where these can be directly tested experimentally.

\section{methods}

The Bi$_{2}$Te$_{3}$ crystal was grown in two steps. Firstly, the 
stoichiometric compound was synthesized starting by heating the pure elements
(Bi and Te) in an ampoule at about 300$^{\circ}$C in hydrogen environment to 
eliminate oxidized species. On suite, the ampoule was evacuated at room temperature and then heated to 200$^{\circ}$C for 24 hours and then up to 587$^{\circ}$C at a rate of 0.5$^{\circ}$C/min and left there for 3 hours. The system was 
then cooled down at 0.2$^{\circ}$C/min in 72 hours. At this point, the obtained
products were checked by X-ray diffraction and the synthesis of the compound
was confirmed. In a second step, the single crystal was grown by heating the 
obtained compound inside an evacuated and sealed quartz ampoule up to 271$^{\circ}$C at 0.5$^{\circ}$C/min and kept in this value for 2 hours. The temperature was then increased to 587$^{\circ}$C at a rate of 2$^{\circ}$C/min and the ampoule was left at this temperature for 24 hours. The
system was then cooled down to room temperature at 0.05$^{\circ}$C/min.
The grown single crystal was then characterized by Laue diffraction. This 
analysis showed a good crystalline quality. In addition, the Seebeck coefficient was measured and a XPS analysis of the sputtered (111) surface was carried out and no impurity was detected.

ARPES measurements of the bulk band dispersion of Bi$_2$Te$_3$ were performed on the SGM-3  beamline of the ASTRID synchrotron radiation facility.\cite{Hoffmann2004} The  Bi$_2$Te$_3$ single crystals have been cleaved in vacuum at room temperature and successively measured at $\sim$ 70~K. ARPES spectra have been acquired at different photon energies, spanning from  14~eV to 32~eV. The crystals have been aligned in two different high symmetry directions: $\bar{\Gamma}$--$\bar{\rm M}$  and $\bar{\Gamma}$--$\bar{\rm K}$. The energy and angular resolution for ARPES measurements were better than 20~meV and 0.2$^{\circ}$, respectively. 

The LDA and $GW$ calculations were performed with the all-electron FLAPW codes 
{\sc fleur}{}~\cite{fleur} and {\sc spex}{},\cite{friedrich2010-spex} respectively. 
We used the experimental lattice structure of Ref.~\onlinecite{Wyckoff1964}. 
For the LDA calculations, we employed an angular momentum cutoff for the muffin-tin 
spheres of $l=10$ and a plane-wave cutoff in the interstitial 
region of 4.5~bohr$^{-1}$. The SOC was incorporated self-consistently employing the
second-variation technique.\cite{li1990} 
We use the same basis for the wave functions in the $GW$ calculations
and, furthermore, employ an angular momentum cutoff of $l=5$ and a 
linear momentum cutoff of 2.9~bohr$^{-1}$ for the representation of the
screened interaction and related quantities.\cite{kotani2002,friedrich2010-spex} 
For the calculation of the bulk band structure in the $\Gamma$--Z--F--$\Gamma$--L path, we have used a 
4$\times$4$\times$4 \kv-point mesh for the screened interaction 
$W$ and evaluated the quasiparticle energies on 190 \kv~points along the path. 
However, for the band structure projected onto the (111) surface (in the
$\bar{\Gamma}$--$\bar{\rm M}$ and $\bar{\Gamma}$--$\bar{\rm K}$ 
directions), we had to calculate self-energy corrections for almost 
2000 extra \kv~points (no interpolation technique was employed). 
In order to save computation time, we resorted to a 2$\times$2$\times$2 
\kv-point set for $W$ in this case. Tests show that the quasiparticle 
energies differ by maximally 50~meV between calculations with a 2$\times$2$\times$2 
and a 6$\times$6$\times$6 sampling, whereas, for most of the \kv~points, 
the differences are much smaller, in particular for the $\Gamma$ and
Z points (0.6 and 14~meV, respectively). 
Therefore, 2$\times$2$\times$2 turns out to give enough quantitative
and qualitative accuracy for the purpose of this work. We have included 500 bands and semicore \emph{d} states of Bi and Te. 
In addition, to have an accurate description of high-lying states and to avoid 
linearization errors,\cite{friedrich2006,friedrich2011,michalicek2013} we have included 
two local orbitals per angular momentum up to $l=3$  for each atom.

In contrast to most $GW$ calculations that include spin-orbit 
interactions, we take the SOC already into account in the reference
system\cite{Sakuma2011} instead of in an \textit{a posteriori} correction, employing
the four-component spinor wave functions. The self-energy
thus acquires terms that couple the two spin channels, enabling a many-body renormalization
of the SOC itself (for a detailed discussion see Ref.~\onlinecite{Aguilera2013-2}).

As in Ref.~\onlinecite{Aguilera2013}, we solve the quasiparticle equation in the basis of 
the LDA single-particle states explicitly. This takes the off-diagonal elements of the 
self-energy  into account, allowing for changes in the quasiparticle wave 
functions, which proved to be critical, for example, to 
describe the highest valence band of \bite~correctly.\cite{Aguilera2013} 
Although still being a one-shot approach, we go beyond the usual
\textit{perturbative} solution of the quasiparticle equation of motion, in which the
quasiparticle wave functions are approximated by the corresponding LDA single-particle states, which 
requires only the diagonal elements of the self-energy to be calculated. In Ref.~\onlinecite{Nechaev2013-2},
it was concluded that studies beyond the perturbative one-shot approach were required for \bite, as the
dependence on the one-particle starting point was found to be stronger than for \bise. 

For a comparison to the ARPES results, the theoretical Fermi level has been adjusted (shifting it up from mid gap by 0.13~eV) so as to match at best the measured upper valence band. This procedure is justified because of the strong $n$-doping of the crystals.

\section{Results and Discussion}

\begin{figure}
\includegraphics[width=8.2cm]{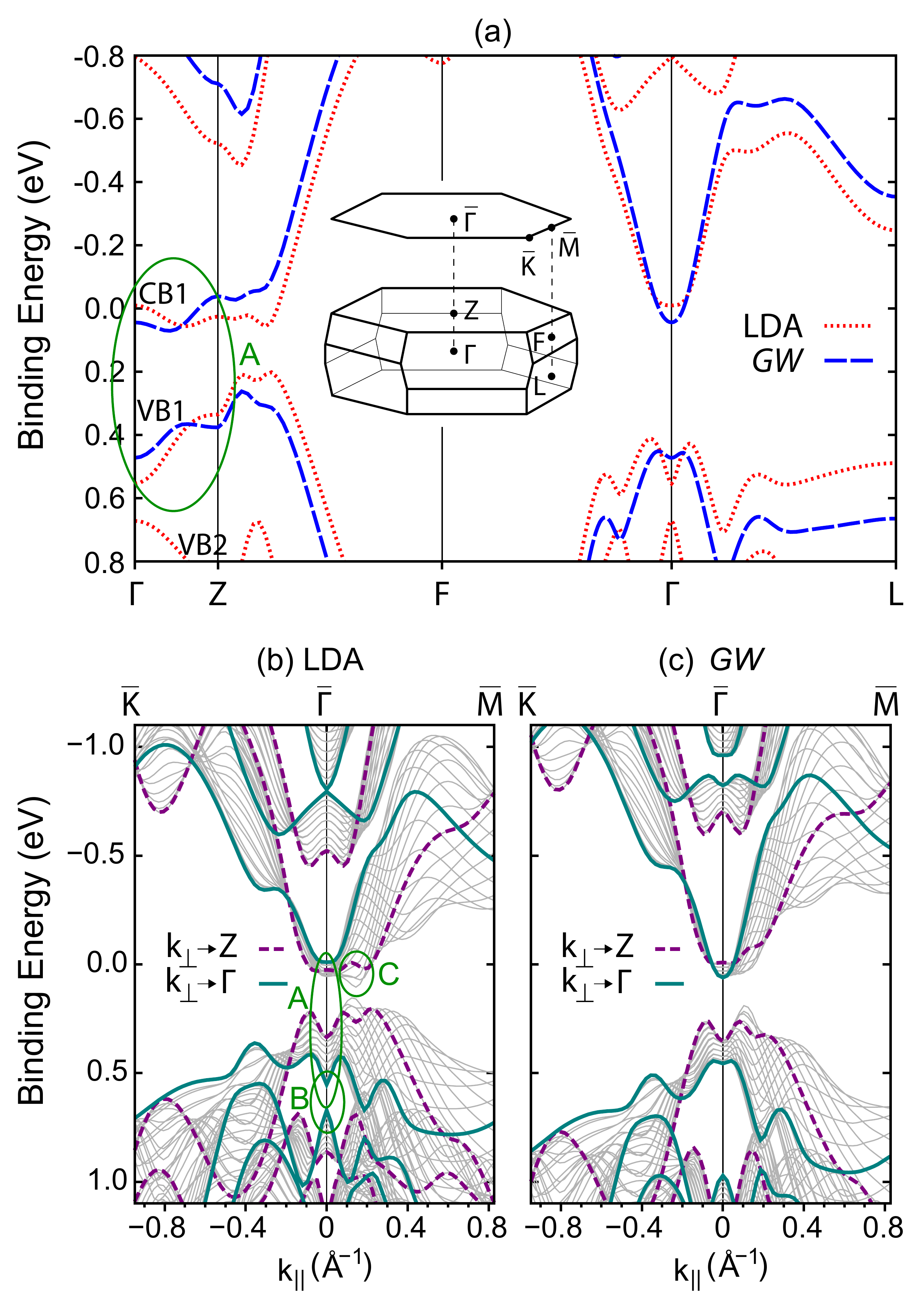}%
\caption{(Color online) (a) Bulk band dispersion of Bi$_2$Te$_3$ calculated by LDA (dotted red line) and $GW$ (long-dashed blue line). VB1, VB2 and CB1 denote the bands for which the dispersion is compared to the experimental data. The inset shows the bulk Brillouin zone and its projection onto the (111) surface. (b) and (c) Bulk bands projected onto the (111) surface along the $\bar\Gamma$--$\bar{\rm K}$ and $\bar\Gamma$--$\bar{\rm M}$ directions calculated with LDA and $GW$, respectively. The darker solid and dashed lines correspond to paths including the $\Gamma$ and Z point, respectively. The green ovals denote the regions \textbf{A}, \textbf{B}, and \textbf{C} that were chosen for a comparison to the experimental data because the differences between LDA and $GW$ are maximal there. The binding energy scales in the calculations have been shifted by 0.13~eV in order to facilitate the comparison with experiment.}  
\label{fig:1}
\end{figure}

The calculated bulk band structure of Bi$_2$Te$_3$ is shown in Fig.~\ref{fig:1}(a) for both the LDA and $GW$ approaches. Figures~\ref{fig:1}(b) (LDA) and (c) ($GW$) show the bulk bands projected onto the (111)-surface BZ along the $\bar\Gamma$--$\bar{\rm K}$ and $\bar\Gamma$--$\bar{\rm M}$ directions. The LDA band structures are in good agreement with previous publications.\cite{Eremeev:2010b,Zhang2009,Yazyev2010,Nechaev2013-2} Note that the band projections in Fig.~\ref{fig:1}(b-c) are not slab calculations of the surface electronic structure. Therefore, they do not show the topological surface state nor any other surface states that might be present in the projected band gaps. 

The LDA calculation [Fig.~\ref{fig:1}(b)] shows an almost-direct fundamental gap of 50~meV with both the VBM and CBM in the 
$\bar\Gamma$--$\bar{\rm M}$ direction. The $GW$ approximation [Fig.~\ref{fig:1}(c)] confirms the position of the 
VBM in $\bar\Gamma$--$\bar{\rm M}$, but places the CBM in the $\Gamma$--Z direction instead, giving an indirect 
band gap of 120~meV, in better agreement with experimental values (130--170~meV, Refs.~\onlinecite{austin1958,li1961,sehr1962,thomas1992}).
The VB exhibits the characteristic M-shaped dispersion (darker solid lines) around \gb~which is symptomatic of the band inversion at~$\Gamma$. 
The transition from LDA to $GW$ results in a flattening of the M-shape dip. As shown by $k\cdot p$ perturbation theory in the case of Bi$_2$Se$_3$ in 
Ref.~\onlinecite{Yazyev2012}, this effect stems from the band inversion due to SOC: 
$GW$ increases the band gap with respect to LDA for all \kv~points (as it happens 
for most trivial gaps) but reduces the gap at $\Gamma$ because it is here inverted.
The inclusion of quasiparticle effects ``flattens'' the band even though the band inversion persists.\cite{Aguilera2013}

\begin{figure*}
\includegraphics[width=1\textwidth]{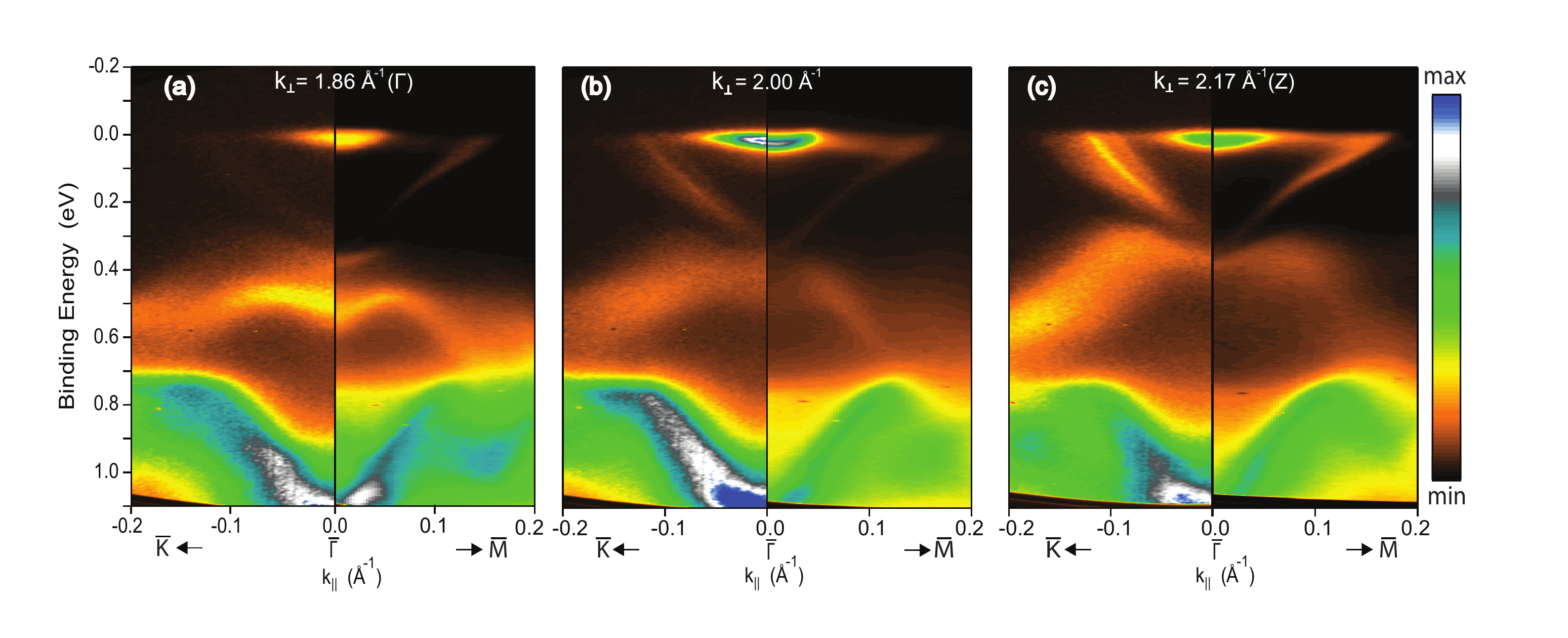}%
\caption{(Color online) (a-c) Photoemission intensity from Bi$_2$Te$_3$(111) along the $\bar{\rm K}$--$\bar\Gamma$ and $\bar\Gamma$--$\bar{\rm M}$ directions for different values of the crystal momentum perpendicular to the surface ${\rm k}_{\rm \perp}$.} 
\label{fig:2}
\end{figure*}

It should be noted that the dispersion of the highest valence band in 
the present $GW$ calculation [Fig.~\ref{fig:1}(a)] and the one by Yazyev \etal~in 
Ref.~\onlinecite{Yazyev2012} differ significantly. In particular, 
the M shape which is flattened--but persists--in our case, 
almost disappears in Ref.~\onlinecite{Yazyev2012}. These discrepancies have been 
recently discussed in Ref.~\onlinecite{Aguilera2013-2} to be due to the \textit{a posteriori} 
inclusion of SOC in Ref.~\onlinecite{Yazyev2012}. Furthermore, we do not
employ a pseudopotential approximation, nor a plasmon-pole model in our
calculations. 

It is interesting to notice that the band structure of \bite~shows pronounced differences with that of the analogue compound \bise.
The \bise~band structure, investigated by ARPES and $GW$, reveals the VBM to be precisely at the $\Gamma$ point.\cite{Nechaev2013} 
The flattening effect is in fact much stronger and restores the convex shape to the upper valence band. Since the CBM is also found at $\Gamma$ this results in the debated direct band gap.
This does not seem to happen in \bite~because of a qualitative difference in the bulk-projected bands:
 In the case of \bite~[Fig.~\ref{fig:1}(b-c)], the highest occupied state at $\bar{\Gamma}$ 
is at the Z point, whereas in the case of \bise~it is placed at the $\Gamma$ point. This implies that the disappearance of the 
M shape is really affecting the highest occupied state in \bise~(and therefore it changes the nature of the fundamental 
band gap) but not in \bite.

In Fig.~\ref{fig:1}(b-c), when LDA 
and $GW$ results are compared, there are three features 
that show larger differences between the two approaches. These three features are shown enclosed by green 
ovals and are labelled \textbf{A}, \textbf{B} and \textbf{C}. \textbf{A} refers to the differences in the 
binding energies at $\Gamma$ and Z and their position with respect to each other, as well as the dispersion 
along the  $\Gamma$--Z direction. \textbf{B} labels the 
region of a projected band gap in the occupied states. This gap is significantly smaller in LDA. 
Finally, \textbf{C} points to the location of a minimum of the conduction band in the $\bar\Gamma$--$\bar{\rm M}$ 
direction in LDA (actually the absolute CBM) which is not present in $GW$. This is also relevant for the discussion 
of the position of the CBM and the nature of the band gap. In the comparison to the ARPES spectra, we focus the discussion on
these three regions. 

Figure~\ref{fig:2} shows parts of a large set of photon-energy-dependent ARPES data, the photoemission intensity along the $\bar{\rm K}$--$\bar{\Gamma}$ and $\bar{\Gamma}$--$\bar{\rm M}$ is shown for three different photon energies. The data have been converted from the raw format (intensity as a function of kinetic energy and emission angle) to the intensity as a function of binding energy and ${\rm k}_\parallel$. The relative intensity of the bands is different for the two measured directions, such that intensity jumps occur at ${\rm k}_{\rm \parallel}=0$ for some binding energies. This is ascribed to polarization-dependent matrix elements in the photoemission process. Several features are immediately identified in these spectra: The conduction band is clearly discernible at the Fermi level and ${\rm k}_{\rm \parallel} \approx 0$. This is expected due to the strong $n$-doping of the samples. The topological surface state with its characteristic Dirac cone shape is visible between the conduction band and the valence band. As expected for a two-dimensional state, it does not disperse with ${\rm k}_{\rm \perp}$. The top of the valence band shows a clear dispersion with photon energy. Finally a very intense state can be seen at high binding energy. This state does not present any dispersion with ${\rm k}_{\rm \perp}$ either.

\begin{figure}[t!]
\includegraphics[width=.5\textwidth]{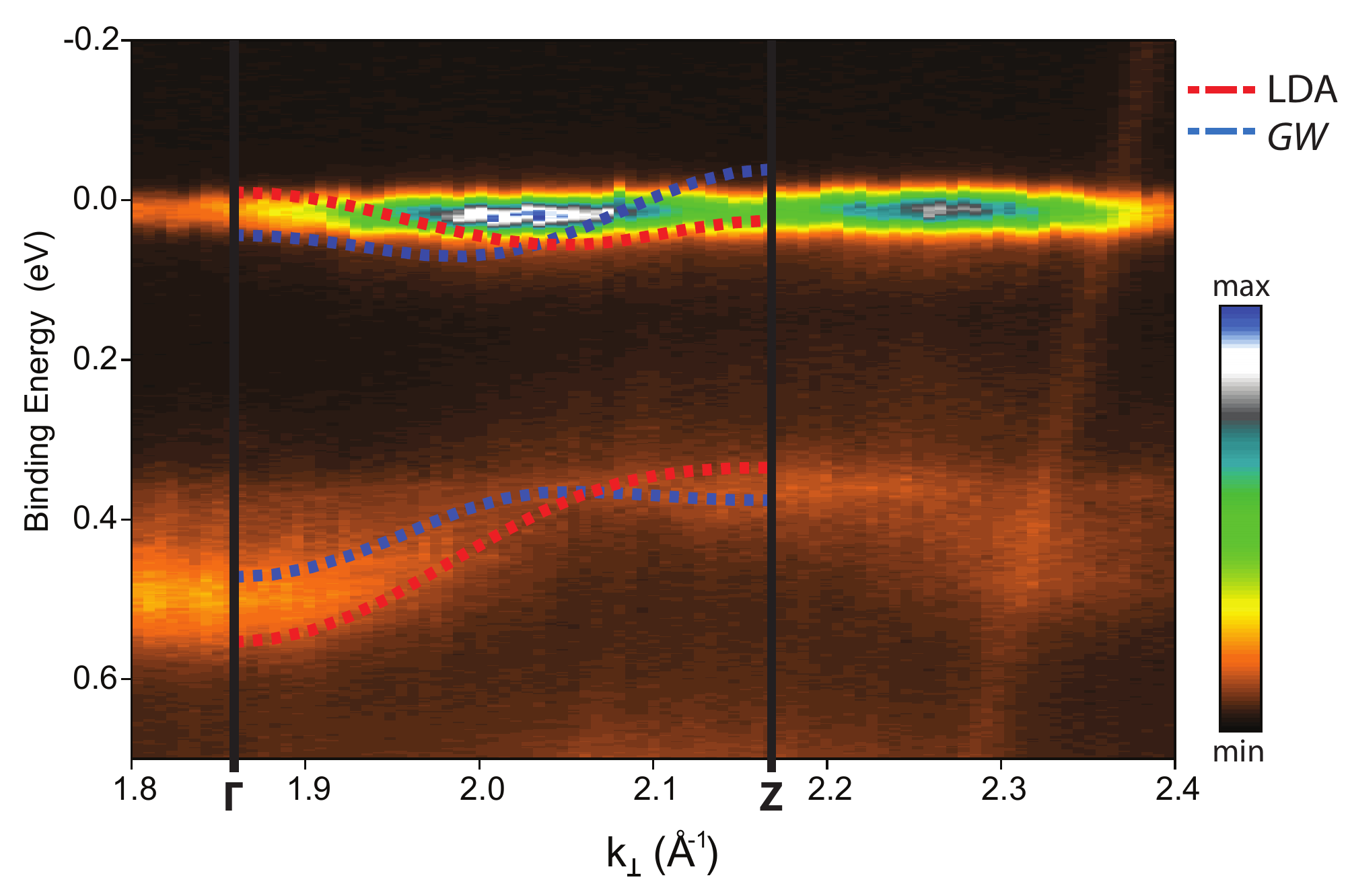}%
\caption{(Color online) Photoemission intensity in normal emission as a function of ${\rm k}_{\rm \perp}$, \ie~the dispersion along the $\Gamma$--Z direction with superimposed bands calculated with the LDA (red dashed line) and the $GW$ approximation (blue dashed line). The position of the bulk $\Gamma$ and Z points is marked.}
\label{fig:3}
\end{figure}

Figure~\ref{fig:3} shows the photoemission intensity in normal emission extracted from a large number of such images as a function of binding energy and ${\rm k}_{\rm \perp}$, the component of the crystal momentum perpendicular to the surface. Note that ${\rm k}_{\rm \perp}$ is not conserved in the photoemission process and that the conversion from photon energy to ${\rm k}_{\rm \perp}$ therefore requires assumptions about the final states in the photoemission process. Here we assume free-electron like final states such that \mbox{$ {\rm k}_{\rm \perp}~=~\sqrt{2m_e/\hbar^2}(V_0+E_{kin} \cos^2(\theta))^{1/2}$} where $\theta$ is the electron's emission angle and V$_0$ is the inner potential.\cite{Himpsel:1980} The inner potential needs to be chosen such that the location of the observed critical points in the dispersion agrees with the expected position along ${\rm k}_{\rm \perp}$, \ie~with the critical points placed either at the $\Gamma$ or at the Z point of the bulk BZ. In the data, a clear dispersion of the top valence band is seen and, guided by the band structure calculation along $\Gamma$--Z, we determine $V_0=1.0$~eV, such that the high binding-energy extremum of the band is placed at the $\Gamma$ point. Note that the dispersion of the top valence band as well as the value of the inner potential is quite different from the case of Bi$_2$Se$_3$ where the binding-energy maximum of the top valence band is found at Z and $V_0$ is approximately 11.8~eV.\cite{Nechaev2013,Xia2009}

In order to judge the importance of many body effects on the band structure of \bite~and the possibly improved description by using the $GW$ approximation, we compare the results of the calculations with the experiment in the three spectral regions \textbf{A}-\textbf{C} where there are clear differences between the LDA and $GW$ results.

\textbf{A--} As mentioned above, comparing the detailed calculated band dispersion to the experimental results is not without problems because ${\rm k}_{\rm \perp}$ is not conserved in the photoemission process and our simple assumption of free electron final states might not be justified. Nevertheless, if we restrict the comparison to normal emission, the observed states lie along the $\Gamma$--Z direction (in the absence of surface umklapp processes), greatly simplifying the analysis. Figure~\ref{fig:3} shows a comparison of the two calculated band dispersions for the $\Gamma$--Z direction and the experimental data in the region of the lowest conduction band CB1 and highest valence band VB1 (see FIG~\ref{fig:3}). 

For VB1, the predicted dispersion in the $\Gamma$--Z shows a binding-energy maximum at $\Gamma$ and a binding-energy minimum at Z and this ordering has also been used to guide the choice of the inner potential $V_0$ for the free electron final state. The main difference between the LDA result and the $GW$ result is that the band width is larger in LDA. When comparing this to the experimental data in Fig.~\ref{fig:3}, it appears that the experimental band width lies in between the LDA and $GW$ results. Another difference between the two theoretical approaches is that the $GW$ dispersion shows a small dip in the vicinity of the Z point, which is not present in LDA. The experimental result does show this small dip in agreement with $GW$. 

For the lowest conduction band CB1, LDA and $GW$ both predict the binding-energy maximum to be located in between the high symmetry points (at ${\rm k}_{\rm \perp}=$2.06~\AA$^{-1}$ and 1.99~\AA$^{-1}$, respectively). It should be noted that the dispersion obtained by a $GW$ calculation where the SOC is included \textit{a posteriori} [see \eg~Figs.~1(b) and 2(b) in Refs.~\onlinecite{Aguilera2013-2} and~\onlinecite{Yazyev2012}, respectively] does not show any binding-energy maximum of CB1 in between $\Gamma$ and Z. As evidenced in Fig.~\ref{fig:3}, the LDA bandwidth is smaller, so that the band  has a noticeably higher binding energy at Z in LDA than in $GW$. A careful inspection of the experimental data appears to confirm the prediction of a binding-energy maximum in between $\Gamma$ and Z, with CB1 apparently reaching a maximum at  ${\rm k}_{\rm \perp}=$2.035~\AA$^{-1}$. Far from this maximum, we can only detect the tail of the CB1 because it disperses above the Fermi level. Since the photoemission matrix elements can have an important role in the spectral intensity profile for different photon energies and CB1 never disperses very clearly below the Fermi level, it remains unclear whether the CB1 band has a higher binding energy at $\Gamma$, in agreement with $GW$, or at Z, as predicted by LDA.

\begin{figure}[t!]
\includegraphics[width=.5\textwidth]{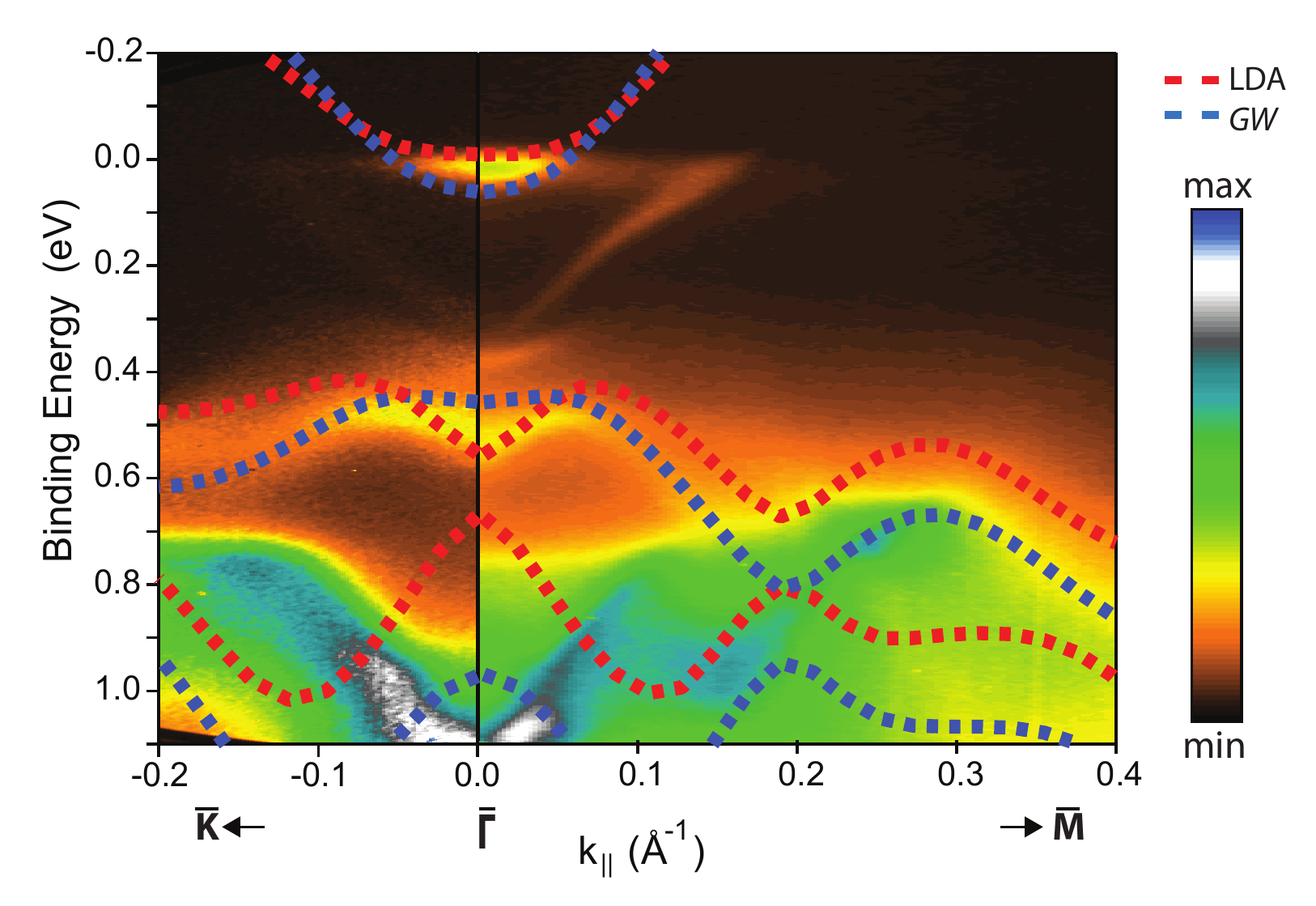}%
\caption{(Color online) (a) Photoemission intensity for ${\rm k}_{\rm \perp}$ near the bulk $\Gamma$ point along the $\bar{\rm K}$--$\bar \Gamma$--$\bar{\rm M}$ direction with the LDA (red dashed line) and $GW$ bands (blue dashed line). }
\label{fig:4}
\end{figure}

\textbf{B--} The spectral region of the projected band gap between VB1 and VB2 near $\bar{\Gamma}$ is distinctly different in the LDA and $GW$ calculations. In LDA, the gap is much narrower at $\bar{\Gamma}$ because of the smaller gap between VB1 and VB2 at the bulk $\Gamma$ point. Further towards $\bar{\rm M}$ and $\bar{\rm K}$, the lower edge of the gap is determined by the VB2 dispersion close to the bulk Z point [dashed lines in Fig.~\ref{fig:1}(b-c)]. Experimentally, one might be able to distinguish between the two scenarios by following the dispersion of the VB1 and VB2 bands. Such a comparison close to the bulk $\Gamma$ point is shown in Fig.~\ref{fig:4}. As in the data of Fig.~\ref{fig:2}, the VB1 is easily identifiable in the spectra  whereas VB2 is not. Instead, one finds a very intense V-shaped feature in the binding-energy region of VB2, as also seen in Fig.~\ref{fig:2}(a-c). We assign this feature to a surface state in the projected band gap between VB1 and VB2. This is confirmed by the fact that the state's dispersion is independent of ${\rm k}_{\rm \perp}$, something that is already apparent in Fig.~\ref{fig:2}(a-c) and confirmed by fitting the dispersion at different photon energies. The existence of this surface state, which cannot be degenerate with a bulk state, implies that the projected band gap at $\bar \Gamma$ must be quite wide, significantly wider than predicted by LDA. 
Surface calculations 
within LDA\cite{Eremeev:2010b,Plucinski2013,Pauly2012} have indicated the existence of a similar V-shaped surface 
state in \bise~and \sbte~but not in \bite.\cite{Eremeev:2010b,Herdt2013} 
For this latter material, the surface states show a wrong dispersion (W-shaped) 
due to the W-shape of the too small LDA bulk projected gap.

The existence and dispersion of the surface state thus strongly favors the 
presence of a larger projected band gap, in better agreement with the $GW$ result.
However, the projected band gap is likely to be even wider than predicted by 
$GW$. The VB2 does not fall within the spectral 
range investigated near $\Gamma$ in Fig.~\ref{fig:4}, but, as 
predicted by $GW$ [dashed line in Fig.~\ref{fig:1}(c)], it can be observed 
around Z [as a blue feature in Fig.~\ref{fig:2}(c)] at \mbox{${\rm k}_{\rm \parallel}\approx$~-0.17~\AA$^{-1}$ }
off-normal emission and approximately 0.9~eV of binding energy. 
While the quasiparticle correction, in fact, yields a considerably
larger gap than LDA, it still seems to underestimate the experimental
one. Self-consistent $GW$ calculations might be able to improve the
agreement with experiment. 

\begin{figure}[h!]
\includegraphics[width=0.5\textwidth]{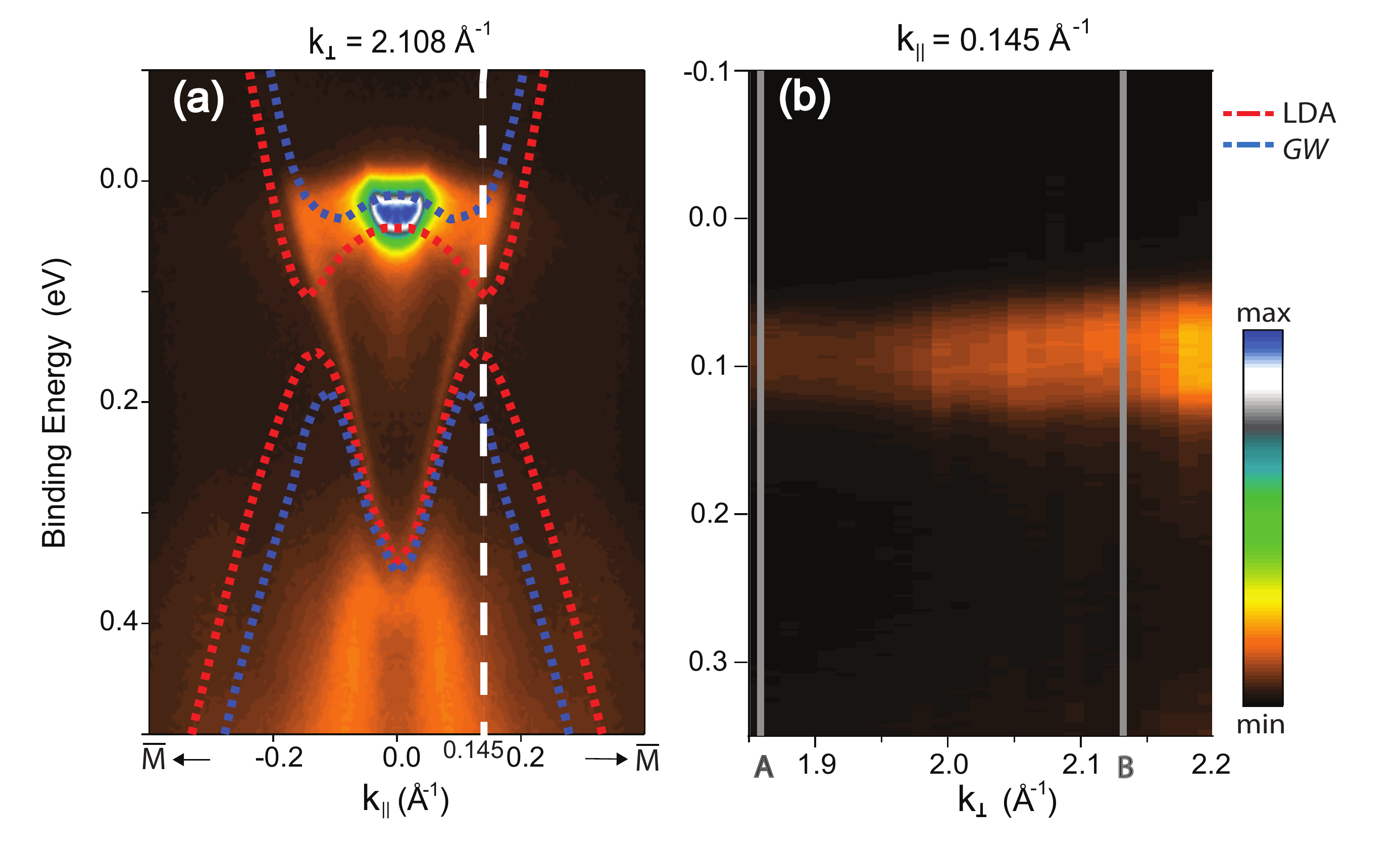}%
\caption{(Color online) (a) Photoemission intensity along the $\bar{\rm M}$--$\bar \Gamma$--$\bar{\rm M}$ direction for the ${\rm k}_{\rm \perp}$ value for which LDA predicts the CBM. LDA and $GW$ bands are shown as red and blue dashed lines, respectively. (b) Photoemission intensity for the ${\rm k}_{\rm \parallel}$ value of the LDA CBM as a function of binding energy and ${\rm k}_{\rm \perp}$. The corresponding ${\rm k}_{\rm \parallel}$ value is marked in (a) by a vertical dashed line. We define here points A and B as the projections of $\Gamma$ and Z on the \mbox{${\rm k}_{\rm \parallel}$ = 0.145~\AA$^{-1}$} line. The only discernible feature is caused by the topological surface state.}
\label{fig:5}
\end{figure}

\textbf{C--} A final distinct difference between the LDA and $GW$ results is the position of the CBM. As seen in Fig.~\ref{fig:1}(b-c), the $GW$ bands show the projected CBM to be placed at $\bar{\Gamma}$ whereas the LDA calculation shows the CB1 dropping below the value at $\bar{\Gamma}$ for ${\rm k}_{\rm \parallel}$ = 0.145~\AA$^{-1}$ along the $\bar \Gamma$--$\bar{\rm M}$ direction with the highest binding energy at ${\rm k}_{\rm \perp}$ = 2.108~\AA$^{-1}$. Such a signature of a CBM off the $\bar \Gamma$ point is never observed in the experiment. To illustrate this, Fig.~\ref{fig:5}(a) shows a spectrum along the $\bar{\rm M}$--$\bar \Gamma$--$\bar{\rm M}$ direction for ${\rm k}_{\rm \perp}$ = 2.108~\AA$^{-1}$. Clearly, the experimental results cannot be reconciled with the CB1 dispersion predicted by LDA while they are more consistent with the $GW$ result. Figure~\ref{fig:5}(b) shows the photoemission intensity in the $\bar \Gamma$--$\bar{\rm M}$ direction at  ${\rm k}_{\rm \parallel}$=0.145~\AA$^{-1}$ as a function of binding energy and ${\rm k}_{\rm \perp}$, i.e. along a line parallel to the $\Gamma$--Z line. The ${\rm k}_{\rm \perp}$ components of the two points marked as A (${\rm k}_{\rm \parallel}$=0.145~\AA$^{-1}$, ${\rm k}_{\rm \perp}$=1.86~\AA$^{-1}$) and B (${\rm k}_{\rm \parallel}$=0.145~\AA$^{-1}$, ${\rm k}_{\rm \perp}$=2.17~\AA$^{-1}$) correspond to those of the $\Gamma$ and Z point, respectively. A single, non-dispersing feature is observed and assigned to the topological surface state. A drop of the CB1 below the Fermi level, as predicted by LDA, is not seen for any value of ${\rm k}_{\rm \perp}$ between A and B. 

The position of the absolute CBM as well as of the VBM in Bi$_2$Te$_3$ is still a subject of debate.\cite{Kioupakis2010,Yavorsky2011,Nechaev2013-2} Our measurements show that the CBM is placed along \mbox{$\Gamma$--Z}. This clearly excludes the possibility of a direct gap since the VBM is off $\bar{\Gamma}$. This can be seen from the fact that the Dirac point of the topological state, which cannot be degenerate with bulk states, is buried between the M-shaped valence band branches which reach a smaller binding energy at ${\rm k}_{\rm \parallel}$ $\neq 0$.
Both LDA and $GW$ agree in placing the VBM along the $\bar{\Gamma}$--$\bar{\rm M}$. According to LDA this will create an almost-direct band gap with the CBM as mentioned before.
In the case of $GW$ the VBM is found nearly in the same position as for LDA but it gives rise to the indirect band gap.
While both approaches predict the VBM along $\bar{\Gamma}$--$\bar{\rm M}$, the experimental result does not confirm this clear directional preference. We find that the energies of the two local binding-energy minima along $\bar{\Gamma}$--$\bar{\rm M}$ and $\bar{\Gamma}$--$\bar{\rm K}$, respectively at ${\rm k}_{\rm \perp}$$\sim$1.97~\AA$^{-1}$ and ${\rm k}_{\rm \perp}$$\sim$2.16~\AA$^{-1}$, are actually very similar.

As mentioned above, the VBM is predicted both in LDA and $GW$ to be placed in the 
$\bar\Gamma$--$\bar{\rm M}$ direction, specifically, for the ${\rm k}_{\rm \perp}$ shown in 
Fig.~\ref{fig:5}(a). However, the binding energy 
of this maximum is somewhat larger in the experiment in better agreement with $GW$ than with LDA. 
The $GW$ calculation  has thus acted again correcting the LDA in the right direction. We would 
like to point out that the widely used \textit{perturbative} one-shot $GW$ approach (\ie~calculating only the 
diagonal elements of the self-energy) has shown to be drastically wrong in this direction of the 
BZ [see Fig.~5(b) of Ref.~\onlinecite{Nechaev2013-2} and its discussion]. This is a clear case of 
unsatisfactory quasiparticle dispersions, which is caused by the neglect of hybridization effects that arise 
from the off-diagonal part of the self-energy. \cite{Aguilera2013} 

\section{conclusions}

We have analyzed the LDA and $GW$ band structures of \bite~and compared them 
to ARPES measurements. In particular, we have analyzed in detail three regions of the spectra in which 
qualitative differences between LDA and $GW$ are observed.

We have also discussed the position of the VBM and CBM and the nature of the gap. 
The LDA calculation shows an almost direct gap of 50~meV with both the VBM and CBM in the
$\bar\Gamma$--$\bar{\rm M}$ direction. The $GW$ approximation confirms the position of the 
VBM along $\bar\Gamma$--$\bar{\rm M}$, 
but places the CBM in the $\Gamma$--Z direction instead, giving an 
indirect band gap of 120~meV, in better agreement with experimental values
(130--170~meV, Refs.~\onlinecite{austin1958,li1961,sehr1962,thomas1992}). Our ARPES results 
 confirm the position of the CBM along $\Gamma$--Z, as predicted by $GW$ and the position of the VBM away from the $\bar{\Gamma}$ point. However, the valence band does not reach noticeably smaller binding energies along $\bar\Gamma$--$\bar{\rm M}$  than along $\bar\Gamma$--$\bar{\rm K}$.

The one-shot $GW$ calculations including the off-diagonal elements of the 
self-energy\cite{Aguilera2013} and a consistent treatment of spin-orbit interactions\cite{Aguilera2013-2} 
constitute a significant overall improvement to the LDA results and produce quasiparticle band structures 
in better agreement with ARPES measurements. This (together with the recent findings\cite{Nechaev2013,Yazyev2012} 
about the direct gap of \bise) emphasizes the importance of many-body effects on the band structure of this 
family of topological insulators.

\begin{acknowledgments}
We thank I. A. Nechaev and G. Bihlmayer for fruitful discussions.  
We gratefully acknowledge financial support by the VILLUM foundation, CNPq and FAPEMIG.     
The theoretical work was supported by the Alexander von Humboldt Foundation through a postdoctoral fellowship, and by the Helmholtz 
Association through the Virtual Institute for Topological Insulators (VITI). 
\end{acknowledgments}




\end{document}